\documentclass[%
 reprint,
unsortedaddress,
 amsmath,amssymb,
 aps,
floatfix,
]{revtex4-1}

\usepackage{graphicx}
\usepackage{dcolumn}
\usepackage{bm}
\usepackage{xcolor}
\usepackage[normalem]{ulem}

\newcommand{\ket}[1]{|#1\rangle}

\begin{document}


\title{Corner states of two-dimensional second-order topological insulators with a chiral symmetry and broken time reversal and charge conjugation.}%

\author{Joseph Poata}
\affiliation{%
School of Chemical and Physical Sciences and MacDiarmid
Institute for Advanced Materials and Nanotechnology, Victoria University of
Wellington, PO Box 600, Wellington 6140, New Zealand}

 \author{Fabio Taddei} 
 \affiliation{NEST, Instituto Nanoscienze-CNR and Scuola Normale Superiore, I-56126, Pisa, Italy}

\author{Michele Governale}
\affiliation{%
School of Chemical and Physical Sciences and MacDiarmid
Institute for Advanced Materials and Nanotechnology, Victoria University of
Wellington, PO Box 600, Wellington 6140, New Zealand}

\date{\today}
\begin{abstract}
Two-dimensional second-order topological insulators are characterized by the presence of topologically protected zero-energy bound states localized at the corners of a flake.
In this paper we theoretically study the occurrence and features of such corner states inside flakes in the shape of a convex polygon.
We consider two different models, both in Cartan class IIIA, the first obeying inversion symmetry and the other obeying a combined $\pi/4$ rotation symmetry and time-reversal symmetry ($\hat{C}_4^z\hat{T}$). By using an analytical effective model
of an edge corresponding to a massive Dirac fermion, we determine the presence of a corner state between two given edges by studying the sign of their induced masses and derive general rules for flakes in the shape of a convex polygon.
In particular, 
we find that the number of corner states in a flake is always two in the first model, while in the second model there are either 0, 2 or 4.
To corroborate our findings, we focus on flakes of specific shapes (a triangle and a square) and use a numerical finite-difference approach to determine the features of the corner states in terms of their probability density. In the case of a triangular flake, we can change the position of corner states by rotating the flake in the first model, while in the second model we can also change their number.
Remarkably, when the induced mass of an edge is zero the corresponding corner state becomes delocalized along the edge.
In the case of a square flake and the model with $\hat{C}_4^z\hat{T}$ symmetry, there is an orientation of the flake with respect to the crystal axes, for which the corner states extend along the whole perimeter of the square.

\end{abstract}
\maketitle

\section{Introduction}
In recent years, there has been a growing interest in topological states of matter. One prominent example is Topological Insulators (TI)\cite{Fu2007,Hasan2010,Liu2010,Qi2011}. These materials are characterised by a gapped band-structure that cannot be adiabatically connected, that is without closing the gap, to the atomic-limit of a trivial band insulator. In TIs, the protection of the topological state is provided by time reversal symmetry. 
A striking consequence of band topology is the \textit{bulk-boundary correspondence}: the $d-1$-dimensional surface of a $d$-dimensional TI hosts gapless excitations described by a massless Dirac Hamiltonian. These excitations are topologically protected from disorder that preserves the time-reversal symmetry of the system. In a two-dimensional TI, the gapless excitations are the helical edge states that give rise to the spin Hall effect\cite{Kane2005,konig2007}. 

Another class of topological materials are Higher-Order Topological Insulators (HOTI)\cite{Xie2021,Benalcazar2017,Benalcazar2017PRB,Song2017,Langbehn2017}.
A $n$-th order TI in $d$ dimensions  obeys a  \textit{generalised bulk-boundary correspondence}: it hosts gapless boundary states in $d-n$ dimensions. For example, a three dimensional second-order TI hosts one-dimensional conducting states at the hinges between different sides.
The general concept is that the effective boundary Hamiltonian of a HOTI is that of a Dirac fermion  with a mass gap which changes sign on different boundaries giving rise to localised mid-gap excitations at the phase boundary\cite{Song2017,Schindler2018,Schindler2018NatPhys,Geier2018,Sheng2019,Schindler2020,Chen2020,Wieder2020,Krishtopenko2021,Roy2021,Sekine2022,Krishtopenko2023}.
A large number of
different models for HOTIs, characterised  by different underlying symmetries and topological invariants, have been proposed\cite{Benalcazar2017,Benalcazar2017PRB,Song2017,Langbehn2017,Schindler2018,Geier2018,Franca2018,vanMiert2018,Khalaf2018,Ezawa2018,Ezawa2018b},
just to mention the earlier works in the literature. 

So far there have been a few candidate systems and materials predicted to be a HOTI, such as twisted bilayer graphene\cite{Park2019}, graphdiyne\cite{Sheng2019}, breathing Kagome and pyrochlore lattices\cite{Ezawa2018}, phosforene\cite{Ezawa2018b}, WTe$_2$ and MoTe$_2$\cite{Wang2019}, antiperovskite materials\cite{Fang2020}, in cubic semiconductor quantum wells\cite{Krishtopenko2021} and induced by a magnetic field in Quantum Spin Hall systems\cite{Langbehn2017,Ezawa2018,Ren2020,Chen2020,Krishtopenko2023}.
Remarkably, a few materials have been observed to behave as a HOTI.
These are bismuth\cite{Schindler2018NatPhys}, 
Bi$_4$Br$_4$\cite{Noguchi2021,Shumiya2022}, and WTe$_2$\cite{Choi2020,Lee2023}.
Furthermore,  HOTIs have been realised with  classical waves in phononic, photonic and circuit systems (see Ref. \cite{Xie2021} for a review).

The existence of the mid-gap boundary states in HOTIs is a consequence of the bulk topology\cite{Schindler2020}. However, the properties of these states depend on the microscopic model of the material and the boundary.
In this paper, we focus on the corner states of second-order two dimensional HOTIs. We study two different models for the bulk HOTI: one that preserves inversion symmetry and the second that preserves the combined symmetry $\hat{C}_4^z\hat{T}$, where $\hat{T}$ is time-reversal and $\hat{C}_4^z$ the rotation by $\pi/4$ about the $z$ axis. Both models belong to Cartan class AIII, i.e. time-reversal and charge-conjugation symmetries are broken but there is an additional  chiral anti-symmetry\cite{Schnyder2008,Ludwig2016}. Our results are specific to this class. 
We consider a flake in the shape of a convex polygon and study the position and features of the corner states.
Our main findings are that, for the model with inversion symmetry, the number of mid-gap corner states is always two, whereas, for the model with $\hat{C}_4^z\hat{T}$ symmetry, the number of corner states can be changed by modifying the shape or orientation of the flake with respect to the crystal axes, and it is possible to completely eliminate the corner states. 
Remarkably, in the case of a square flake, for a particular orientation with respect to the crystal axes the corner states extend along
the whole perimeter of the square.
Finally, a circular flake can be seen as the limit of a convex regular polygon when the number of sides goes to infinity. For both models, there are well-localised (corner) states on the circumference at positions determined by the zeros of the induced mass.

The paper is organised as follows: In Section \ref{sec:Model}, we introduce the models of the two-dimensional HOTIs considered here. In Section \ref{sec:linear-edge} we derive the effective Hamiltonian for an arbitrary linear edge and in \ref{sec:corner-states}  we discuss the condition for the formation of a zero-energy state localised at the corner between two adjacent edges. In Section \ref{sec:results} we compare the predictions of the analytical model of Section \ref{sec:corner-states} with the results obtained by solving the quantum-confinement problem by means of a finite-difference method. Finally in Section \ref{sec:conclusions} we will draw some conclusions.

\section{Model and Formalism}
\label{sec:Model}
In this section we introduce two continuous model for a two-dimensional HOTI.
We describe the two-dimensional HOTI by the Hamiltonian $H=H_{\text{TI}}+H_{\text{M,I(II)}}$. The Hamiltonian $H_{\text{TI}}$ describes a TI 
\begin{align}
\label{eq:H_TI}
H_{\text{TI}}= & m(\hat{\mathbf{k}})\sigma_0\tau_z 
+ A (\hat{k}_x\sigma_x+\hat{k}_y\sigma_y)\tau_x ,
\end{align}
where $\hbar\hat{\mathbf{k}}$ is the momentum operator, $\tau_i$ is the $i$-th Pauli matrix in orbital space and $\sigma_i$ the $i$-th Pauli matrix in spin-space. 
For convenience we have defined 
\begin{align}
m(\hat{\mathbf{k}})=m_0+m_2 \hat{\mathbf{k}}^2.   
\end{align}
For the sake of definiteness, we assume  $m_0<0$, $m_2>0$, and $A>0$. 
The Hamiltonian $H_{\text{TI}}$ supports gapless edge states. These edge states are gapped by the term $H_{\text{M,I(II)}}$. We consider two models 
\begin{align}
H_{\text{M,I}}=&
\frac{C}{\sqrt{2}}\left(\sigma_x+\sigma_y\right)\tau_0\quad\quad\text{model I},  \\    
H_{\text{M,II}}=& B(\hat{k}^2_y-\hat{k}^2_x)\sigma_0\tau_y \quad\quad\text{model II}, 
\end{align}
with $B>0$ and $C>0$. 
The Hamiltonians of models I and II belong to Cartan class AIII. Time reversal and charge conjugation symmetry are broken while there is a chiral anti-symmetry, represented by the operator $\sigma_z \tau_x$, that is $\left\{H, \sigma_z \tau_x\right\}=0$, with $\{\dots \}$ denoting the anticommutator. All the results derived in the following pertain to a 2D HOTI in this Cartan class\cite{Schnyder2008,Ludwig2016} .
Model I  has inversion symmetry and describes the presence of a Zeemann field \cite{Langbehn2017,Schindler2020} in the direction $\frac{1}{\sqrt{2}}(\hat{\mathbf{x}}+\hat{\mathbf{y}})$. 
Here, inversion symmetry implies $\hat{\mathcal{I}}H(\mathbf{k})\hat{\mathcal{I}}^\dagger=H(-\mathbf{k})$, with $\hat{\mathcal{I}}=\sigma_0\tau_z$.
Model II has been introduced in Ref.~\cite{Schindler2018} and has the combined symmetry $\hat{C}_4^z\hat{T}$, where $\hat{T}$ is the time-reversal operator and $\hat{C}_4^z$ the rotation operator by $\pi/4$ along the $z$ axis. 
We have chosen model I due to its simplicity and the fact that can be implemented in a spin Hall system with an in-plane magnetic field, and model II as it is a very common model in this field.
\subsection{Linear edge}
\label{sec:linear-edge}
\begin{figure}
    \centering
\includegraphics[width=0.8\columnwidth]{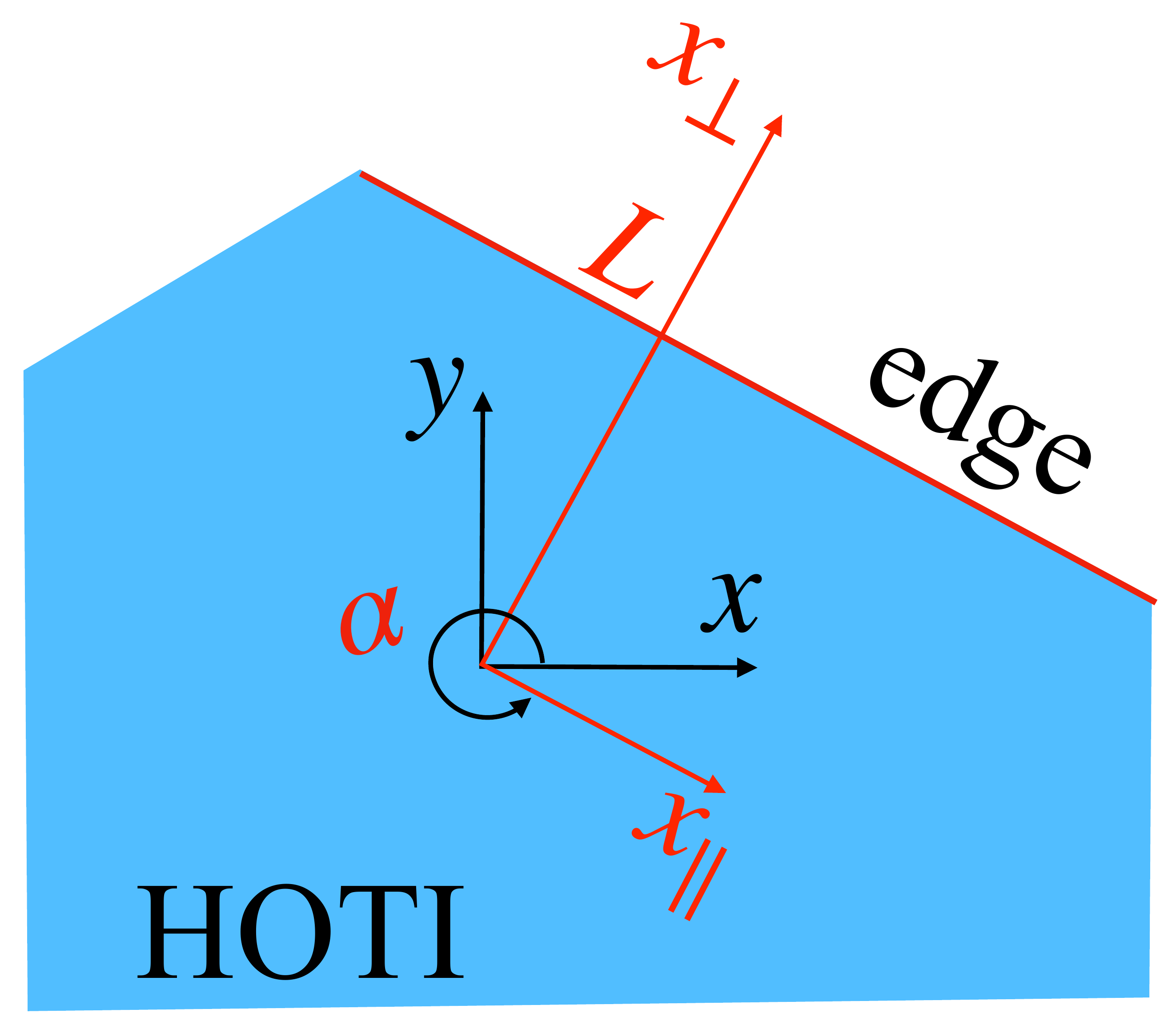}\\
    \caption{Schematic description of a linear edge (red) and of the rotated coordinate system $(x_\parallel,x_\perp)$. The edge is located at $x_\perp=L$, as shown. The rotation angle $\alpha$ defines the direction of the edge. }
    \label{fig:edge}
\end{figure}

In this subsection, we derive the effective Hamiltonian for a generic linear edge. We consider a linear edge at a distance $L$ from the coordinate origin. The direction of the edge with respect to the crystal is defined by the angle $\alpha$ by which we need to rotate the coordinate axes, so that the new coordinates, $x_\parallel$ and  $x_\perp$ are, respectively, parallel to the edge and perpendicular to it pointing outwards (away from the region with the HOTI). A schematic description of the edge and of the rotated coordinate system is shown in Fig.~\ref{fig:edge}. 
We take $\alpha$ in the interval $[0,2\pi)$. 
The coordinate transformation is described by
\begin{align}
\left(
\begin{array}{c}
x\\
y
\end{array}
\right) 
=
\left(
\begin{array}{cc}
\cos(\alpha) & -\sin(\alpha)\\
\sin(\alpha) & \cos(\alpha)
\end{array}
\right)
  \left(
\begin{array}{c}
x_\parallel\\
x_\perp
\end{array}
\right).  
\end{align}
It is important to notice that this transformation affects only the spatial coordinates and leaves the spinor basis unchanged. This implies that in the different coordinate systems corresponding to different edges (different values of $\alpha$), the meaning of the spinor components remains the same. 
In the new coordinate system, the Hamiltonians read 
\begin{subequations}
\label{HamRot}
\begin{align}
\nonumber
H_{\text{TI}} =&    m(\hat{k}_\parallel, \hat{k}_\perp)\sigma_0\tau_z +A\left[\hat{k}_\parallel (\cos(\alpha)\sigma_x+\sin(\alpha)\sigma_y)\right. \\ 
&\left.  + \hat{k}_\perp (-\sin(\alpha)\sigma_x+\cos(\alpha)\sigma_y)\right]\tau_x\\
H_{\text{M,I}}=&
\frac{C}{\sqrt{2}}\left(\sigma_x+\sigma_y\right)\tau_0\\    
H_{\text{M,II}}=&B\left[\cos(2\alpha)(\hat{k}^2_\perp-\hat{k}^2_\parallel)+2\sin(2\alpha)\hat{k}_\parallel\hat{k}_\perp \right]\sigma_0\tau_y.
\end{align}
\end{subequations}

We consider the semi-plane $x_\perp\in (-\infty,L]$, i.e. with boundary at $x_\perp=L$. We start by taking $H_{\text{M,I(II)}}=0$ and $\hat{k}_\parallel=0$. In this case the edge states are at zero energy and read:
\begin{align}
    |\Phi_{\alpha,1}\rangle=& \frac{ \rho(x_\perp-L)}{\sqrt{2}}\,  \left( \ket{+,\uparrow}- e^{i\alpha} \ket{-,\downarrow}\right)\, ,\\
     |\Phi_{\alpha,2}\rangle=& \frac{\rho(x_\perp-L)}{\sqrt{2}}\,\left( \ket{+,\downarrow} + e^{-i\alpha} \ket{-,\uparrow} \right)\, ,
\end{align}
where we have introduced the basis $\{\ket{\tau,\sigma}\}$, with $\tau\in\{+,-\}$ and $\sigma\in\{\uparrow,\downarrow\}$.
The function $\rho$ defining the transverse profile of the edge states is 
\begin{align}
\label{eq:rho}
\rho(x)=\frac{1}{N}\left( e^{\lambda_+ x}-e^{\lambda_- x} \right)    
\end{align}
with $N$ being a normalisation factor and 
\begin{align}
\lambda_{\pm}=\frac{A\pm\sqrt{A^2+4 m_0 m_2} }{2 m_2}. 
\end{align}
For the sake of simplicity, we assume that the parameters are such that $\lambda_{+}=\lambda_{-}^{*}=k+ i q$ and $k>0$ (similar results can be obtained when both $\lambda_\pm$ are real).

Now, we include first-order linear terms in $\hat{k}_\parallel$ and $H_{\text{M,I(II)}}$ in Eqs.~(\ref{HamRot}) as a perturbation. 
Computing the matrix elements of the perturbation on the basis $\left\{\ket{\Phi_{\alpha,1}},  \ket{\Phi_{\alpha,2}}\right\}$ we obtain the effective Hamiltonian for the surface states
\begin{align}
\label{eq:H_alpha_eff}
H_{\alpha,\text{eff}}=\left(
\begin{array}{cc}
- A \hat{k}_\parallel     & i M_{\text{I,(II)}}(\alpha) e^{-i \alpha} \\
-i M_{\text{I,(II)}}(\alpha) e^{i\alpha}     & A \hat{k}_\parallel 
\end{array}
\right),
\end{align}
where the mass term for the two models are given by
\begin{align}
\label{eq:MI}
M_{\text{I}}(\alpha)=&C \sin(\alpha-\pi/4) \\
\label{eq:MII}
M_{\text{II}}(\alpha)
= & -B\cos(2 \alpha) (k^2+q^2)
\end{align}
The Hamiltonian Eq.~(\ref{eq:H_alpha_eff}) describes massive Dirac fermions, with a mass term that depends on the direction of the edge, $\alpha$. 

We now study the character of the states for $k_\parallel=0$.
To this end, we diagonalise $H_{\alpha,\text{eff}}$ for $k_\parallel=0$ and we obtain the energies $+M_{\text{I}(II)}$ and $-M_{\text{I}(II)}$. 
The corresponding eigenstates, written in the basis $\{ \ket{\tau,\sigma} \}$, are given by
\begin{align}
\ket{v_{+}(\alpha)}&=\frac{1}{2}\left( \ket{+,\uparrow} -e^{i\alpha} \ket{-,\downarrow}-i e^{i\alpha}  \ket{+,\downarrow}-i\ket{-,\uparrow}\right),\\
\ket{v_{-}(\alpha)}&=\frac{1}{2}\left( \ket{+,\uparrow} -e^{i\alpha} \ket{-,\downarrow}+i e^{i\alpha}  \ket{+,\downarrow}+i\ket{-,\uparrow}\right).
\end{align}
\subsection{Corner states}
\label{sec:corner-states}
We now investigate the possible formation of the topological state at the crossing of two edges defined by angles $\alpha_1$ and $\alpha_2$. To be specific, we want to find out whether there is a gap inversion between the Hamiltonians $H_{\alpha_1,\text{eff}}$ and $H_{\alpha_2,\text{eff}}$. In order to do so, it is important to compare the eigenstates.  
We notice that 
the eigenstates relative to $\pm M_{I(II)}(\alpha_1)$ and $\mp M_{I(II)}(\alpha_2)$ are orthogonal, namely
\begin{align*}
\langle{v_{+}(\alpha_1)}\ket{v_{-}(\alpha_2)} &= \langle v_{-}(\alpha_1)\ket{ v_{+}(\alpha_2)} =0.
\end{align*}
On the contrary, the eigenstates relative to $\pm M_{I(II)}(\alpha_1)$ and $\pm M_{I(II)}(\alpha_2)$ overlap as
\begin{align*}
\langle v_{+}(\alpha_1)\ket{v_{+}(\alpha_2)}&=\langle v_{-}(\alpha_1)\ket{v_{-}(\alpha_2)} =\frac{1}{2}\left(1+e^{i (\alpha_2-\alpha_1)}\right).
\end{align*}
Therefore, we can expect a corner state to form where these two states meet if their masses $ M_{I(II)}(\alpha_1)$ and $ M_{I(II)}(\alpha_2)$ have different signs.
We can conclude that a band inversion occurs if 
\begin{align}
\label{eq:corner-condition}
\text{sign}[M_{\text{I}(II)}(\alpha_1)]\ne \text{sign}[M_{\text{I}(II)}(\alpha_2)].
\end{align}
Remarkably, the form of Eq.~(\ref{eq:H_alpha_eff}) implies another important piece of information, namely
the corner states decay differently along the two adjacent edges. In fact, the decay length along an edge characterised by the angle $\alpha$ is simply 
\begin{align}
\label{eq:R_alpha}
R_\alpha=\left|\frac{A}{M_{\text{I(II)}}(\alpha)}\right|.   
\end{align}
In particular, if $M_{\text{I(II)}}(\alpha)\rightarrow 0$ the corner state is no longer localised in the corner but it extends along the edge with vanishing mass. 
\section{Results}
\label{sec:results}
We use the condition Eq.~(\ref{eq:corner-condition}) to predict the number and positions of corner states for a HOTI flake with the shape of a convex polygon.
We also solve the confinement problem numerically by means of a finite-difference method and compare the analytical predictions to the full numerical results. 

We express energies in units of $|m_0|$ and lengths in units of $R_0=A/|m_0|$. To apply the finite-difference method, we introduce a square discretization grid with lattice constant $a$. We find that accurate results are found for $a=0.1 R_0$.  
For the numerical calculations we use the following values of the parameters 
\begin{align*}
 m_2\,\frac{|m_0|}{A^2}=11/40,\quad     B\,\frac{|m_0|}{A^2}=3/4 \quad \frac{C}{|m_0|}=3/4.
\end{align*}

Let us now 
discuss what happens when we change the direction of a given side $i$ of a polygon, while keeping the directions of the other sides fixed.
Since one or both adjacent sides need to change length (note that we are not performing a rotation of the polygon), the polygon will change shape. 
In particular, we are focusing on the transformation $\alpha_i^{\text{old}}\rightarrow \alpha_i^{\text{new}}$ in which the
mass associated to the side $i$ changes sign, i.e.  $\text{sign}(M_{\text{I(II)}}(\alpha_i^{\text{new}}))\ne \text{sign}(M_{\text{I(II)}}(\alpha_i^{\text{old}}))$.
First, we consider the case where the two adjacent sides have masses of equal sign, namely $\text{sign}(M_{\text{I(II)}}(\alpha_{i-1}))= \text{sign}(M_{\text{I(II)}}(\alpha_{i+1}))$.
If before changing $\alpha_i$ there were no corner states at the two ends of side $i$, then after the change there are two corner states. 
Vice versa, if initially there were two corner states, after the change there are no corner states.
The second case is when the two adjacent sides have masses of opposite sign, $\text{sign}(M_{\text{I(II)}}(\alpha_{i-1}))\ne \text{sign}(M_{\text{I(II)}}(\alpha_{i+1}))$. Let's assume that $\text{sign}(M_{\text{I(II)}}(\alpha_{i-1}))\ne \text{sign}(M_{\text{I(II)}}(\alpha_i^{\text{old}}))$. In this case, before the change there is a corner state at the junction between the sides $i-1$ and $i$ and no corner state at the junction between the sides $i$ and $i+1$. After the change, the corner state moves to the  junction between the sides $i$ and $i+1$. In general, we can conclude that by \textit{changing the direction of one side of a polygon the number of corner states changes by 0,-2, or +2.} 
In a similar fashion, it can be shown that in general \textit{the number of corner states is always even.}
\subsection{Model I (inversion symmetry)}
In this section we deduce some general properties of corner states in a convex polygon based on Eqs.~(\ref{eq:corner-condition}) and (\ref{eq:MI}):
\begin{itemize}
    \item[I(a)] \emph{The number of corner states is always two. }
\end{itemize}
The maximum number of corner states in a convex polygon is determined by the number of zero of the induced mass $M_{\text{I(II)}}(\alpha)$ in the interval $[0,2\pi)$. In the case of $M_{\text{I}}(\alpha)$,   Eq.~(\ref{eq:MI}) has only two zeros for $\alpha\in[0,2\pi)$, namely at $\pi/4$ and $(5/4)\pi$,
and  it is positive when $\pi/4<\alpha<(5/4)\pi$.
From this follows that no convex polygon can have all sides with the same sign of the induced mass. 

\begin{figure}
    \begin{center}
\flushleft{(a)}\\
\centerline{\includegraphics[width=0.8\columnwidth]{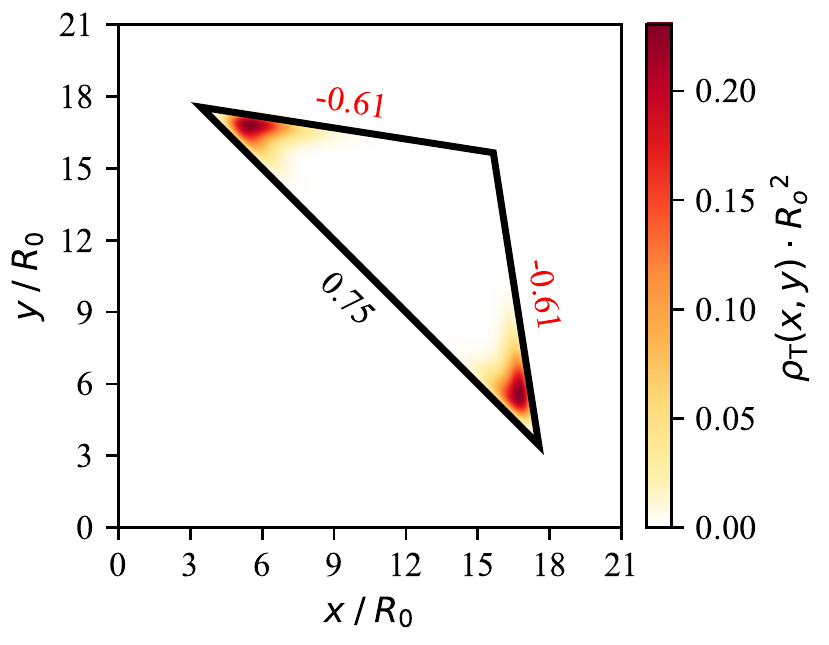}}
\vspace{-0.8cm}
\flushleft{(b)}\\
\centerline{\includegraphics[width=0.8\columnwidth]{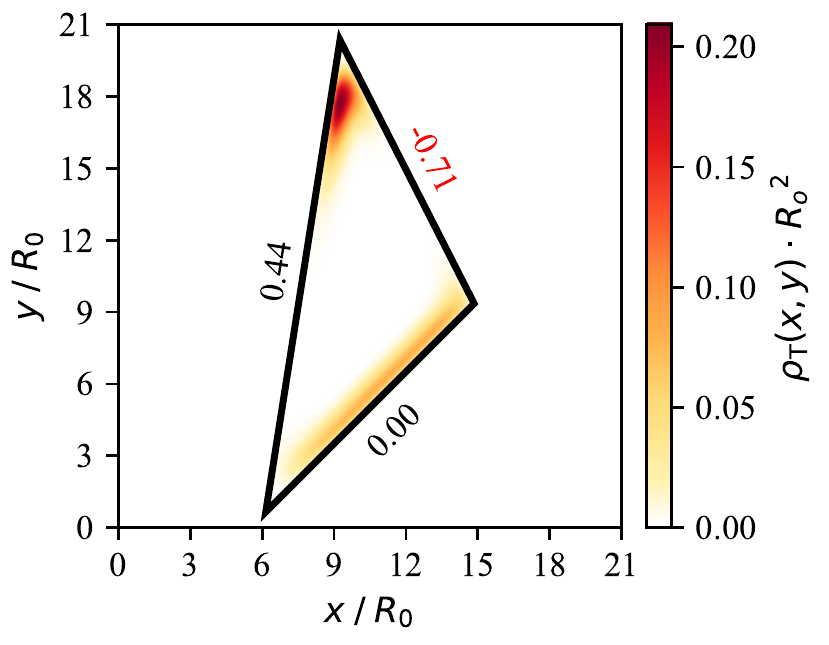}}
\vspace{-0.8cm}
\flushleft{(c)}\\
\centerline{\includegraphics[width=0.8\columnwidth]{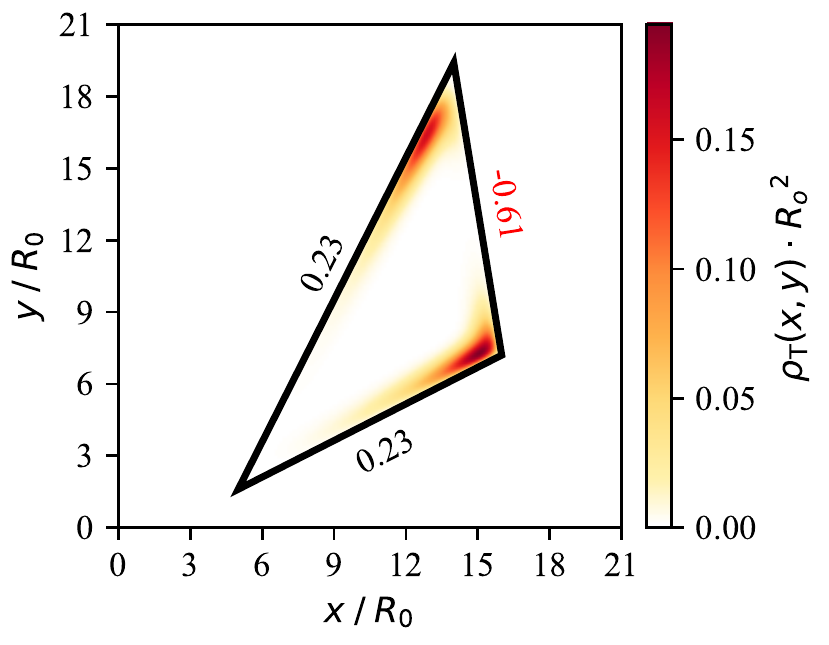}}
\vspace{-0.8cm}
\end{center}
    \caption{We show the total probability density, $\rho_{\text{T}}$, for the two mid-gap states of a flake described by model I, with the shape of an isosceles triangle of base of length $20 R_0$ and base angles of $\pi/5$. The three panels correspond to different orientations with respect to the crystal axes. The outline of the flake is indicated by a solid line. The edges have been annotated with the corresponding values of $M_{\text{I}}$ in units of $|m_0|$ (negative values are in red for ease of reading). For the orientation of panel (a), the two corner states have energies $\approx \pm 10^{-5} |m_0|$ and are well localised. For the orientation of panel (b), the two mid-gap states have energies $\approx \pm 7\times 10^{-4} |m_0|$. However, as $M_{\text{I}}=0$ on one of the edges, the corresponding corner state extends along the entire edge.  For the orientation of panel (c) the two mid-gap states have changed location and  have energies $\approx \pm 4.7\times 10^{-4} |m_0|$.}
    \label{fig:triangleI}
\end{figure}

\begin{figure}
\begin{center}
\flushleft{(a)}\\
\centerline{\includegraphics[width=0.8\columnwidth]{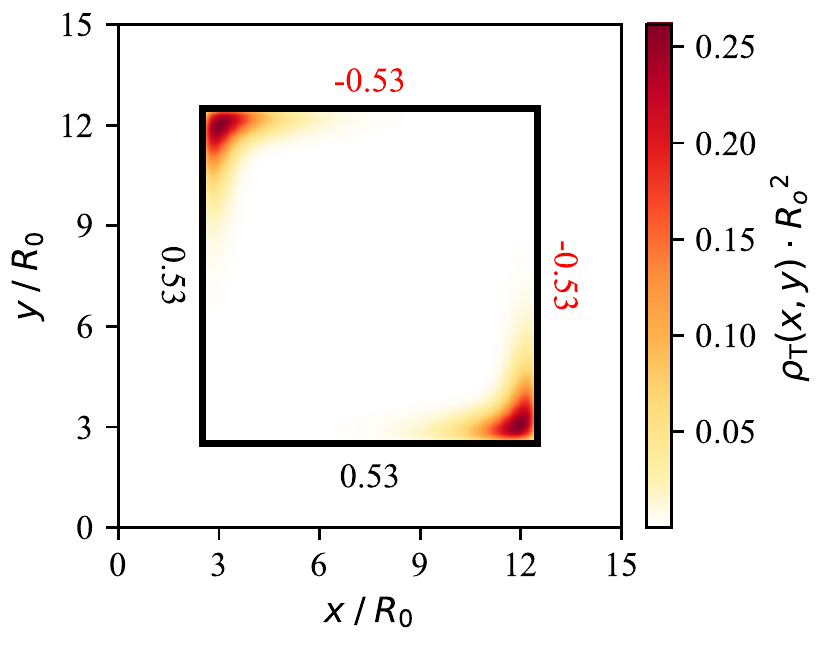}}
\vspace{-0.5cm}
\flushleft{(b)}\\
\centerline{\includegraphics[width=0.8\columnwidth]{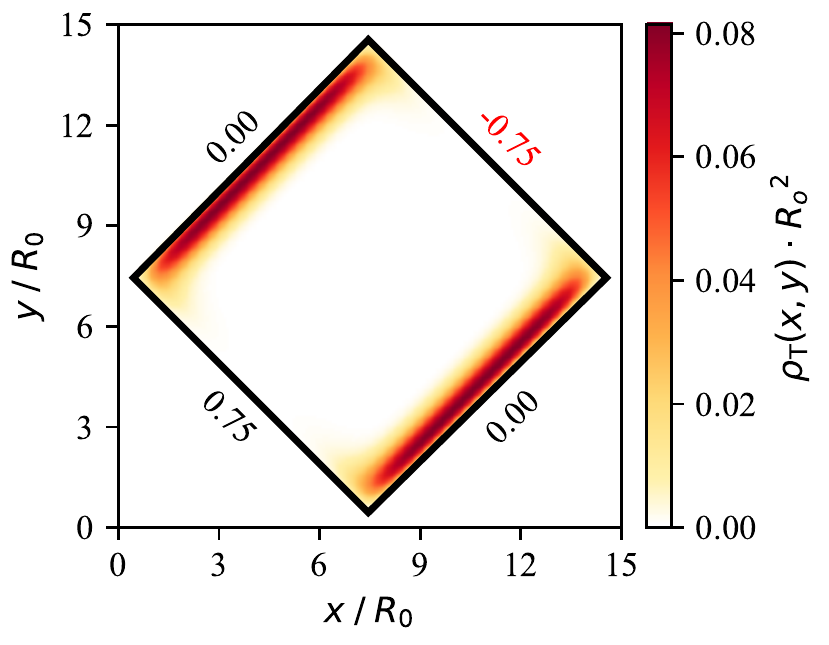}}
\end{center}
        \caption{We show the total probability density, $\rho_{\text{T}}$, for the two mid-gap states of a square flake of side $10 R_0$ described by model I. The two panels correspond to two possible orientations with respect to the crystal axes. The outline of the flake is indicated by a solid line. The edges have been annotated with the corresponding values of $M_{\text{I}}$ in units of $|m_0|$ (negative values are in red for ease of reading). For the orientation of panel (a), the two corner states have energies $\approx \pm 5\times 10^{-5} |m_0|$ and are well localised. For the orientation of panel (b), the two mid-gap states have energies $\approx \pm 1.2\times 10^{-4} |m_0|$. However, as $M_{\text{I}}=0$ on two of the edges, the corresponding corner states extends along the entire edge.   }
    \label{fig:rectangleI}
\end{figure}

\begin{figure}
\flushleft{(a)}\\
\centerline{\includegraphics[width=0.8\columnwidth]{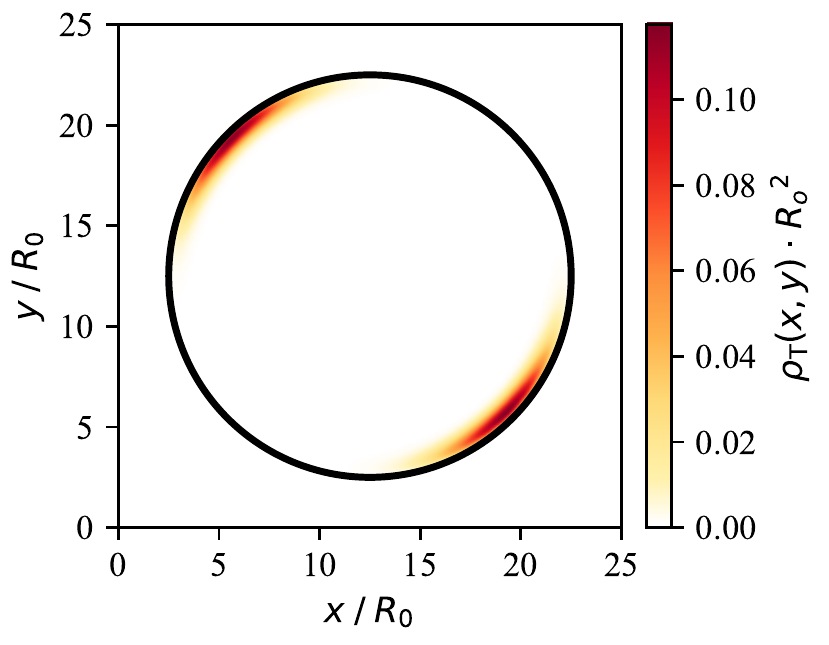}}
\vspace{-0.5cm}
\flushleft{(b)} \\
\centerline{\includegraphics[width=0.8\columnwidth]{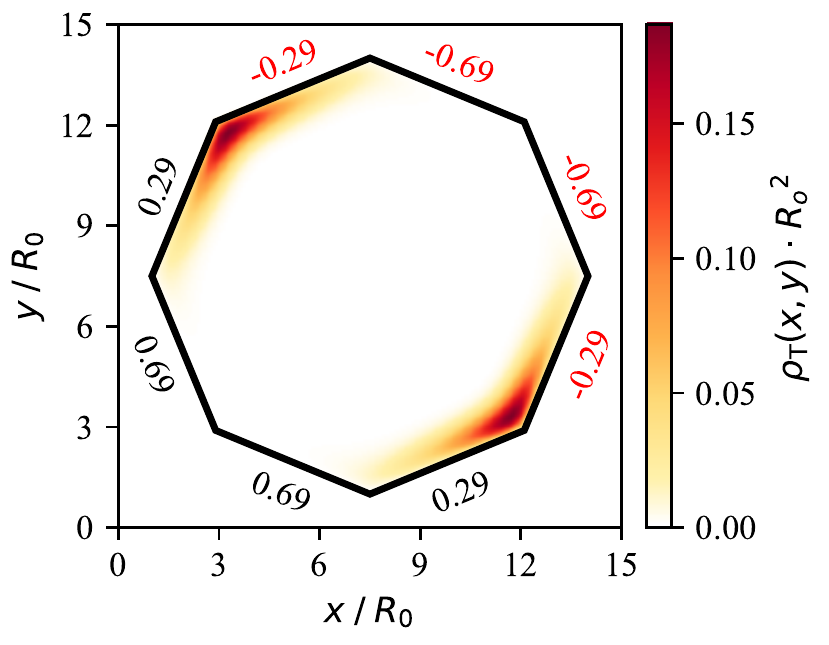}}
         \caption{We show the total probability density, $\rho_{\text{T}}$ for the two mid-gap states of a flake described by model I, with the shape of circle of radius $10 R_0$ (a) and of an octagon of side $10(1+\frac{\sqrt{2}}{2})^{-\frac{1}{2}} R_0$ (b). The outline of the flake is indicated by a solid line. The edges have been annotated with the corresponding values of $M_{\text{I}}$ in units of $|m_0|$ (negative values are in red for ease of reading). }
    \label{fig:circleI}
\end{figure}

In Figs.~\ref{fig:triangleI} and \ref{fig:rectangleI} we show the density plots of the total probability density $\rho_{\rm T}(x,y)$ for the two lowest (close to zero) mid-gap states, for flakes of two different shapes.
Fig.~\ref{fig:triangleI} refers to three triangular flakes of the same shape and dimensions but with different orientations with respect to the crystal axes.
The annotation in the figure indicates the value of the induced mass along the different edges (negative values in red). 
In the flake of Fig.~\ref{fig:triangleI}(a), there are two mid-gap corner states as shown by the probability density.
By slightly rotating clockwise the triangle (not shown), one still finds the two corner states located at the angles at the base, but with different values of the induced masses. 
When the rotation reaches the situation described in Fig.~\ref{fig:triangleI}(b), the induced mass vanishes on one edge and the corresponding corner state extends along the length of the entire edge. 
Increasing the rotation angle further as shown in Fig.~\ref{fig:triangleI}(c), the induced mass changes sign on one edge and the two corner states will then change positions, one is located at one base angle and the other at the vertex angle. In Fig.~\ref{fig:triangleI}(c), it is particularly evident that the corner states decay with a different decay length along the different  edges. We have extracted the decay lengths for the different edges from the total probability density computed numerically and we find an excellent agreement with Eq.~(\ref{eq:R_alpha}). 
In Fig. \ref{fig:rectangleI}, we show results for two square flakes of the same dimension but with different orientations with respect to the crystal axes.
In Fig.~\ref{fig:rectangleI}(a), where the sides are parallel to the axes, the probability density is concentrated at the top-left and bottom-right corners.
Analogously to the case of a triangle, a clockwise rotation of the square by an angle less than $\pi/4$ will not change much (the corner states will still remain located at the two corners). 
When the rotation angle is $\pi/4$, we obtain the special case of Fig. \ref{fig:rectangleI}(b)
where the induced mass $M_{\text{I}}=0$ on two edges. In this case the two topological states extend along the edges with zero mass.  
\begin{itemize}
    \item[I(b)] \emph{In a circle there are only two localised surface states located along the directions $\frac{1}{\sqrt{2}}(-\hat{\mathbf{x}}+\hat{\mathbf{y}})$ and $\frac{1}{\sqrt{2}}(-\hat{\mathbf{x}}-\hat{\mathbf{y}})$.}
\end{itemize}
In Fig.~\ref{fig:circleI}(a), we show the total probability density for the two mid-gap states of a circle. The position of the corner states is determined by the zeros of $M_{\text{I}}(\alpha)$ in Eq.~(\ref{eq:MI}). The circle can be seen as the limit of a regular polygon when the number of sides goes to infinity. To elucidate this point, in  Fig.~\ref{fig:circleI}(b), we show results for an octagonal flake.  
\subsection{Model II ($\hat{C}_4^z\hat{T}$ symmetry)}
We now repeat the analysis of the previous section for model II. The results are obviously different as the functional form of $M_{\text{II}}(\alpha)$ is different than $M_{\text{I}}(\alpha)$.
We now discuss some general properties.

\begin{itemize}
    \item[II(a)] \emph{The number of corner state can be 0, 2, or 4. }
\end{itemize}
The maximum number of corner states in this case is determined by the number of zeros of $M_{\text{II}}(\alpha)$ in the interval $[0,2\pi)$, which is equal to four.   

\begin{figure}
\flushleft{(a)}
\centerline{\includegraphics[width=0.9\columnwidth]{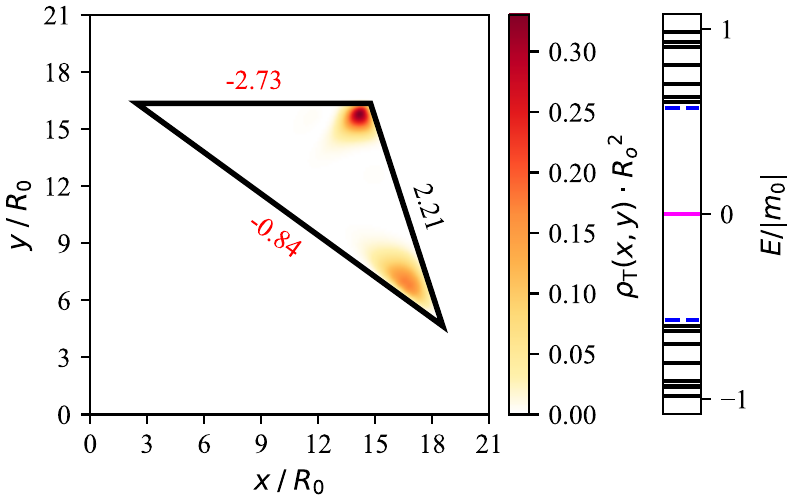}}
\vspace{-0.5cm}
\flushleft{(b)}
\centerline{\includegraphics[width=0.9\columnwidth]{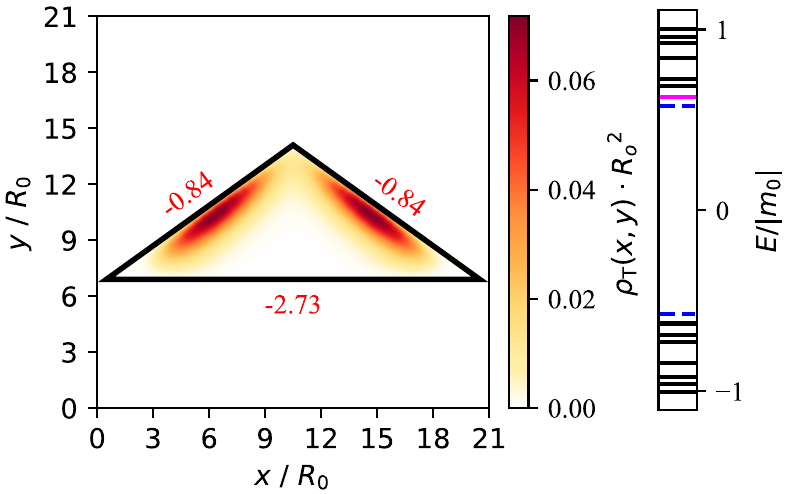}}
  \caption{We show the total probability density, $\rho_{\text{T}}$, for the two corner states (a) and for the four states with lowest absolute value of energy (b) of a flake described by model II, with the shape of an isosceles triangle of base of length $20 R_0$ and base angles of $\pi/5$. The two panels correspond to two possible orientations with respect to the crystal axes. The outline of the flake is indicated by a solid line. The edges have been annotated with the corresponding values of $M_{\text{II}}$ in units of $|m_0|$ (negative values are in red for ease of reading). For the orientation of panel (a), there are two corner states, as predicted by the mass inversion,  which have energies $\approx \pm 5.2 \times 10^{-4} |m_0|$ and are well localised. For the orientation of panel (b), there are no mid-gap states. The four states closest to the zero energy have energies $\approx \pm 0.625 |m_0|$.
 On the right-hand side of each panel, we show the energy spectrum of the flake. The energy of the levels corresponding to the probability density in the main plots, is shown by a solid magenta line. 
 The bulk gap for the Hamiltonian $H_{\text{TI}}$ is given by $\text{min}\left[ |m_0|, \sqrt{-\frac{A^2}{m_2}\left(m_0+\frac{1}{4}\frac{A^2}{m_2}\right)} \right]$ if the square root is real and by $|m_0|$ otherwise. For the parameters used in the numerical simulation, the bulk gap is equal to $\approx 0.575\, |m_0|$ and is indicated by a blue dashed line. 
   } 
    \label{fig:triangleII}
\end{figure}

\begin{figure}
    \centering
\flushleft{(a)}\\
\centerline{\includegraphics[width=0.8\columnwidth]{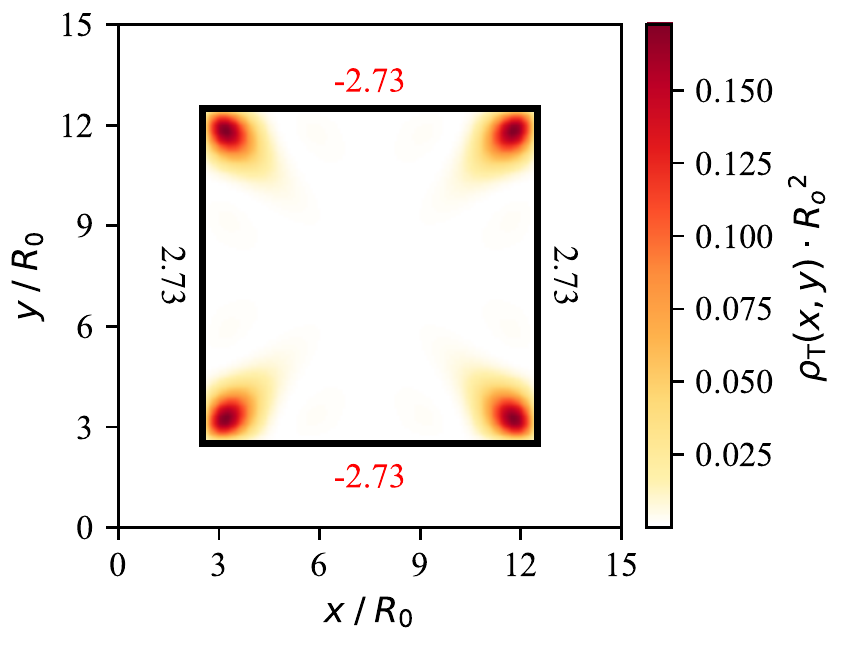}}
\vspace{-0.5cm}
\flushleft{(b)} \\
\centerline{\includegraphics[width=0.8\columnwidth]{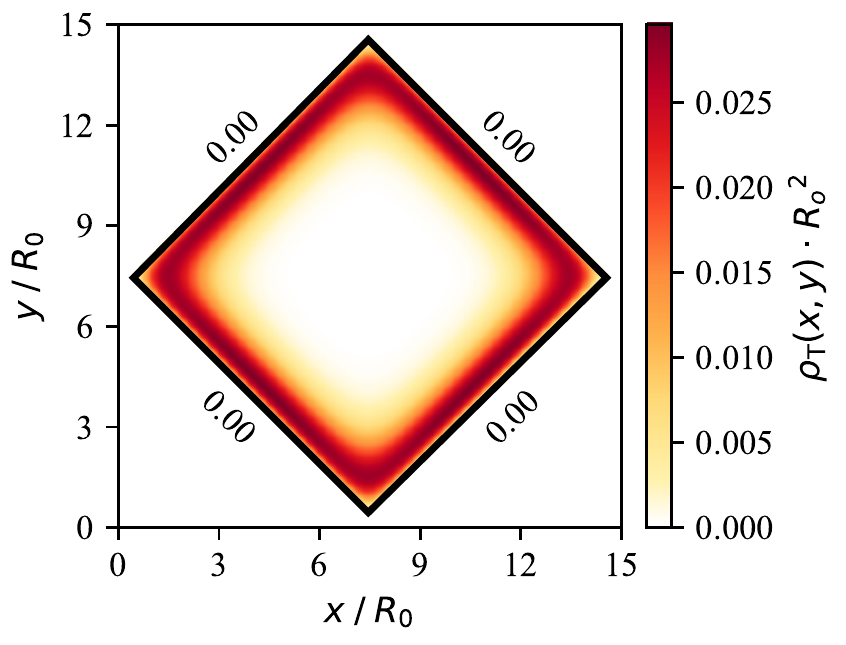}}
      \caption{We show the total probability density, $\rho_{\text{T}}$, for the four mid-gap states of a square flake of side $10 R_0$ described by model II. The two panels correspond to two possible orientations with respect to the crystal axes. The outline of the flake is indicated by a solid line. The edges have been annotated with the corresponding values of $M_{\text{II}}$ in units of $|m_0|$ (negative values are in red for ease of reading). For the orientation of panel (a), the four corner states have energies $\approx \pm 4.8\times 10^{-4} |m_0|$ and are well localised. For the orientation of panel (b), the four mid-gap states have energies $\approx \pm 4.7\times 10^{-2} |m_0|$. However, as $M_{\text{II}}=0$ on all the edges, the corner states extend along the entire perimeter.   }
    \label{fig:rectangleII}
\end{figure}

For this model of HOTI it is possible to find a triangular flake that has zero corner states. In Fig.~\ref{fig:triangleII}, we show results for two triangular flakes of same shape and dimensions but different orientations with respect to the crystal axis. In the flake of Fig.~\ref{fig:triangleII}(a), there are two, well-localized, mid-gap corner states. The probability density associated with these states decays differently along the various edges of the system, 
in agreement with Eq.~(\ref{eq:R_alpha}). On the other hand, for the flake of Fig.~\ref{fig:triangleII}(b), there are no mid-gap corner states (as predicted by the analytical model: the induced masses of all edges are positive). 
\begin{itemize}
        \item[II(b)] \emph{In a rectangular flake there are always four corner states.}
\end{itemize}
In Fig.~\ref{fig:rectangleII}, we show results for two  square flakes. In the total probability density in Fig.~\ref{fig:rectangleII}(a), we clearly see the four corner states.
Notice that $\rho_{\rm T}(x,y)$ is concentrated in the corners, but faint tails are present and extend towards the center of the square.
In general, a rotation of the square only leads to slight changes in the probability distribution $\rho_{\rm T}(x,y)$: the tails do not continue to point toward the center, instead they align along the directions $\frac{1}{\sqrt{2}}(\pm\hat{\mathbf{x}}+\pm\hat{\mathbf{y}})$ (not shown). In Fig.~\ref{fig:rectangleII}(b), we show the case when the mass term $M_{\text{II}}$ vanishes on all edges. The four mid-gap states extend over the entire edge of the flake, effectively forming an extended edge state.

\begin{figure}
\centerline{\includegraphics[width=0.8\columnwidth]{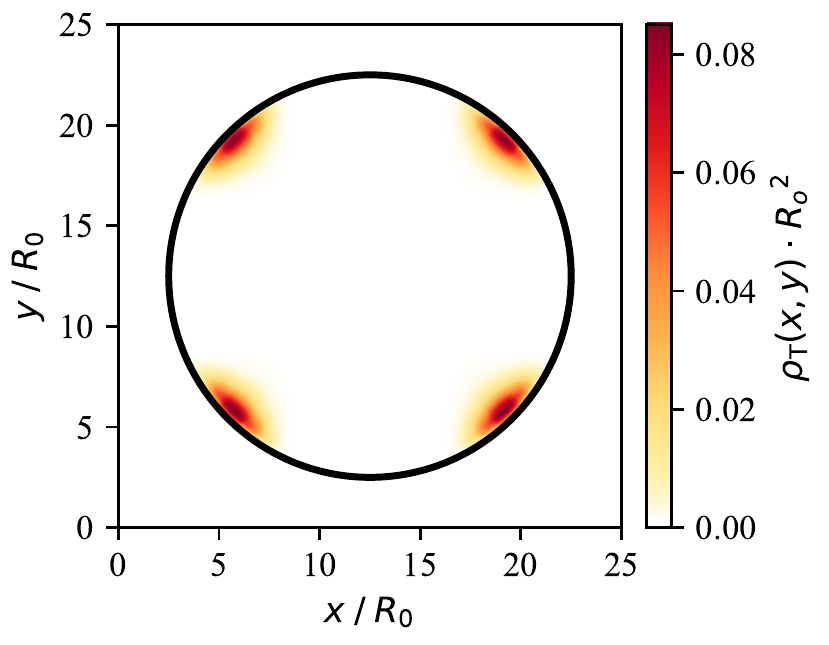}}
 \caption{We show the total probability density $\rho_{\text{T}}$ for the four mid-gap states of a flake described by model II, with the shape of circle of radius $10 R_0$.
 The outline of the flake is indicated by a solid line.
 }
\label{fig:circleII}
\end{figure}

\begin{itemize}    
\item[II(c)] \emph{In a circle there are four localised surface states located along the directions $\frac{1}{\sqrt{2}}(\pm\hat{\mathbf{x}}+\pm\hat{\mathbf{y}})$.}
\end{itemize}
In Fig.~\ref{fig:circleII},
we show the total probability density for the four mid-gap states of a circle. The position of the corner states is determined by the zeros of $M_{\text{II}}(\alpha)$ in Eq.~(\ref{eq:MII}).
\section{Conclusions}
\label{sec:conclusions}
We have presented a study of the corner states in a flake of HOTI of convex polygonal shape.
The Hamiltonian of the second-order TI in Cartan class AIII has been constructed by starting from a Hamiltonian that describes a TI and adding to it an additional term to induce a gap in its edge states. 
We have considered two different gapping Hamiltonians:  
the first one has inversion symmetry, while the second one obeys the combined symmetry $\hat{C}_4^z\hat{T}$.
The results depend on the functional dependence of the induced mass, $M_{\text{I,(II)}}$, on the angle $\alpha$ and they are specific for the particular gapping Hamiltonians considered here.
The model with inversion symmetry shows always two mid-gap (zero energy) corner states, irrespective of the shape (as long as it is convex) of the flake. For special orientations of the flake, when the induced mass vanishes along a side, the corresponding corner state extends along the entire side. 
The model with the combined symmetry $\hat{C}_4^z\hat{T}$ exhibits a richer behaviour, with the number of zero-energy corner states in a flake being zero, two, or four. 
\textcolor{blue}{The case of zero corner states does not imply an exception to the generalised bulk-boundary correspondence, since the shape of the flake does not possess the same symmetry as the crystalline symmetry that protects the second-order topological phase.}  
In a rectangular flake made of a HOTI with the combined symmetry $\hat{C}_4^z\hat{T}$ there are always four corner states. It is also possible to choose the orientation of the flake such that the induced mass vanishes on all sides. In this special case, the four corner states are fused together to form a state which extends along the entire perimeter of the flake -- something unexpected for a second-order TI. Finally for both models, a circular flake shows localised (corner) states on the circumference at positions determined by the zeros of the induced mass.
\bibliography{hoti}{}
\end{document}